\documentclass[twocolumn,preprintnumbers,amsmath,amssymb,superscriptaddress,prl,showpacs]{revtex4}
\usepackage[dvips]{graphicx}
\usepackage{dcolumn}
\usepackage{bm}
\usepackage{colordvi}
\usepackage[dvips]{epsfig}

\usepackage{amssymb}

\begin{document}

\title{Interplay between a hydrodynamic instability and a phase transition: the Faraday instability in dispersions of rodlike colloids}

\author{Pierre Ballesta}
\affiliation{IFF, Institut Weiche Materie, Forschungszentrum J\"ulich, D-52425 J\"ulich, GERMANY.}
\author{Minne Paul Lettinga}
\affiliation{IFF, Institut Weiche Materie, Forschungszentrum J\"ulich, D-52425 J\"ulich, GERMANY.}
\author{S\'ebastien Manneville}
\affiliation{Universit\'e de Lyon, Laboratoire de Physique, Ecole Normale Sup\'erieure de
Lyon, CNRS UMR 5672, 46 All\'ee d'Italie, 69364 Lyon cedex 07, France \& Institut Universitaire de France.}
\date{\today}

\begin{abstract}
Strong effects of the Faraday instability on suspensions of rodlike colloidal particles are reported through measurements of the critical acceleration and of the surface wave amplitude. We show that the transition to parametrically excited surface waves displays discontinuous and hysteretic features. This subcritical behaviour is attributed to the shear-thinning properties of our colloidal suspensions thanks to a phenomenological model based on rheological data under large amplitude oscillatory shear. Birefringence measurements provide direct evidence that Faraday waves induce local nematic ordering of the rodlike colloids. While local alignment simply follows the surface oscillations for dilute, isotropic suspensions, permanent nematic patches are generated by surface waves in samples close to the isotropic-to-nematic transition and above the transition large domains align in the flow direction. This strong coupling between the fluid microstructure and a hydrodynamic instability is confirmed by numerical computations based on the microstructural response of rodlike viruses in shear flow.
\end{abstract}
\maketitle

\section{Introduction}

Recently, the Faraday instability, {\it i.e.} the parametric instability of a fluid layer submitted to vertical vibrations \cite{Faraday:1831}, has appeared as a useful experiment to probe the interplay between the microstructure of various complex fluids and a classical hydrodynamic instability \cite{Kumar:1999, Raynal:1999, Wagner:1999, Lioubashevski:1999, Merkt:2004, Huber:2005, Ballesta:2005, Kityk:2006, Ballesta:2006, Cabeza:2007, Epstein:2010, Hernandez:2010}. Most previous experimental works have focused on the effects of viscoelasticity on the instability threshold \cite{Raynal:1999, Ballesta:2005, Ballesta:2006}, on the subharmonic vs. harmonic response of the surface waves \cite{Wagner:1999, Cabeza:2007}, and on the influence of the microstructure on the surface wave pattern \cite{Kityk:2006, Cabeza:2007}. In shear-thinning semidilute solutions of polymers and of surfactant wormlike micelles, these effects are generally rather small perturbations to the classical instability of Newtonian fluids. However, much stronger effects were found in a few other complex materials, namely clay suspensions \cite{Lioubashevski:1999}, shear-thickening cornstarch suspensions \cite{Merkt:2004}, and dilute shear-thickening wormlike micelles \cite{Epstein:2010}, where localized hysteretic finger-like structures, persistent holes, and high-amplitude strip waves were respectively reported. These original observations were qualitatively interpreted \cite{Lioubashevski:1999, Merkt:2004} and quantitatively modelled \cite{Epstein:2010} as a strong feedback of the surface waves on the fluid microstructure: the oscillatory shear generated by small-amplitude perturbations triggers large variations of the shear rate-dependent viscosity that lead to discontinuous transitions to large-amplitude waves.

Still, to the best of our knowledge, no direct experimental evidence for strong structural modifications induced by Faraday waves has been reported. The aim of the present study is to address this issue in suspensions of {\it fd} virus, a rodlike colloid that presents liquid crystalline phases \cite{Lapointe:1973, Dogic:1997,Dogic:2000}. These viruses are known to easily align under minute external perturbations and the local organization of their microstructure can be simply probed by visualizing the birefringence field \cite{Graf:1993,Lettinga:2004}. Here, three suspensions of {\it fd} viruses at various concentrations, respectively deep into the isotropic domain, close to the isotropic--nematic transition, and in the nematic phase, are submitted to vertical vibrations. After a brief description of the samples and of the experimental set up, we first report on surface wave amplitude measurements that present a strong hysteresis at the onset of the Faraday instability. This subcritical behaviour is accounted for by a simple model based on the shear-thinning properties of our rodlike colloids. We then move to birefringence experiments that reveal local nematic ordering in vibrated suspensions that are initially isotropic at rest. These measurements are successfully compared to numerical simulations based on a Smoluchowski approach~\cite{Dhont:2003a,Lonetti:2008}. Our results point to the existence of an out-of-equilibrium phase transition induced by a hydrodynamic instability in suspensions of rodlike colloids.

\section{Materials and methods}
Our working fluids are suspensions of {\it fd} viruses in a buffer of 20~mM tris-HCl at pH=8.15 as described in Ref.~\cite{Lettinga:2004}. The {\it fd} bacteriophage is a nearly perfectly monodisperse rodlike colloid of length $880$~nm, diameter $6.6$~nm, and persistence length $2.2$~$\mu$m. The number of elementary charges per unit length is around $1$~e$^{-}$.nm$^{-1}$ at pH=8.15. In order to screen the electrostatic interaction between viruses, NaCl is added to the buffer to obtain an ionic concentration of 20~mM. This system presents various phases when the concentration $c$ of {\it fd} virus is increased: isotropic, cholesteric, and smectic \cite{Dogic:1997,Dogic:2000}. Because the free energy between the cholesteric phase and a nematic state is very low, we shall simply refer to the cholesteric phase as nematic in the following. We consider suspensions either in the isotropic or in the nematic phase by focusing on three different samples: an isotropic suspension of concentration $c=5.8$~mg$/$mL well below the isotropic--nematic (I--N) phase transition, an isotropic suspension of concentration $c=11.3$ mg$/$mL just below the I--N phase transition, and a nematic sample of concentration $c=13.6$ mg$/$mL. When submitted to an external shear, {\it fd} viruses tend to align in the shear direction. Deep into the isotropic phase, this alignment causes a dramatic decrease of the viscosity (shear-thinning behaviour) \cite{Graf:1993}, while close to and above the I--N transition, more complex shear-induced phenomena such as vorticity banding, tumbling, and wagging have been reported \cite{Lettinga:2004}.

\begin{figure}
\centering
\scalebox{1}{\includegraphics{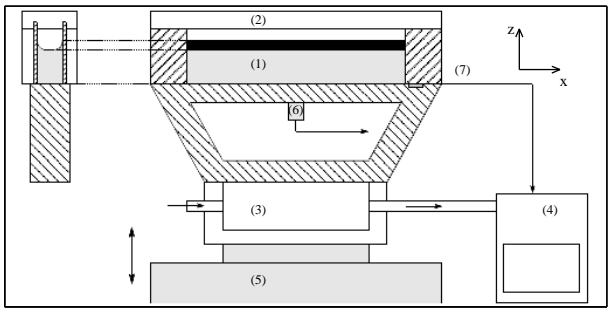}}
\caption{Schematic of the set up. A parallelepipedic cell (1) sealed by a plexiglass lid (2) and thermostated by water circulation (3) from a thermostated bath (4) is vertically vibrated by an electromagnetic shaker (5). The amplitude of the oscillation is recorded via an accelerometer (6), while the temperature of the cell is measured with a probe (7) and feedbacked to the water bath. Left: side view of the cell.}
\label{fig_setup} 
\end{figure}

A dedicated set up (see fig.~\ref{fig_setup}) was developed in order to measure both the birefringence intensity and the amplitude of surface waves in a vertically vibrated layer of {\it fd} suspension. A parallelepipedic glass cell of length 72~mm, thickness 6~mm, and height 8~mm (respectively corresponding to the $x$, $y$, and $z$ directions) is filled up to a height $h=6$~mm with the suspension and vibrated by an electromagnetic shaker (Ling Dynamic Systems V406). 

The driving acceleration is sinusoidal with amplitude $a=10$--100~m.s$^{-2}$ and frequency $f=50$--200~Hz. The suspension is thermostated at $T=20\pm0.5^{\circ}$C by a water circulation beneath the cell. The cell is sealed by a PVC cover to prevent evaporation and surface contamination. It is placed between two polarizers and lit from the back. A fast CCD camera (Mikrotron MC1310) captures the transmitted light at $\sim 1000$~fps. For each value of the acceleration $a$ and frequency $f$, we wait for two minutes so that a steady state is reached, after which two movies are recorded over twenty driving periods ($T=20/f$): (i) a first movie with aligned polarizers from which the amplitude $\xi(x,t)$ is easily extracted as a function of the horizontal position $x$ and time $t$ and (ii) a second movie under crossed polarizers in order to measure the transmitted intensity field $I(x,z,t)$ and to visualize the flow-induced birefringence.$^\dag$ We checked that at the frequencies under study, the rather large aspect ratio of our experimental cell $72/6\gtrsim 10$ always enforces a one-dimensional surface wave pattern along the $x$ direction. In all cases, the surface response was observed to be subharmonic so that $\xi(x,t)$ is well described by:
\begin{equation}
\xi(x,t) =\sin (k x) \sum_{n=0}^{\infty} \xi_n \cos \left(\left(n+\frac{1}{2}\right) \omega t +\psi_n \right),
\end{equation}
where $k$ is the wave number, $\omega=2\pi f$, and $\xi_n$ (resp. $\psi_n$) is the amplitude (resp. phase) of the $n^{\rm\tiny th}$ harmonic. In our experiments $\xi_0\gg \xi_{n>0}$, so that we will only consider $n=0$ in what follows. 

\section{Hysteresis and subcriticality of the Faraday waves}

Following classical protocols for detecting the onset of Faraday waves, we first determine the critical acceleration $a_c$, {\it i.e.} the acceleration above which parametric waves first appear at the fluid surface at a given driving frequency $f$, by (i) increasing the acceleration $a$ by $\delta a/a=0.1\%$ every two minutes and noting the acceleration $a_c^{\rm up}$ when the surface first destabilizes, (ii) waiting for two more minutes for the surface to be fully destabilized and (iii) decreasing the acceleration by $\delta a$ every two minutes until the surface becomes flat again, which defines a second acceleration $a_c^{\rm down}$. A striking difference with experiments in Newtonian fluids is the presence of a large hysteresis in the instability onset: in all three {\it fd} suspensions, $a_c^{\rm up}$ is about $4\%$ greater than $a_c^{\rm down}$. For comparison, in a Newtonian fluid of similar viscosity, no hysteresis is detected, {\it i.e.} $a_c^{\rm up}=a_c^{\rm down}$ to within our precision of $0.1\%$. 

\begin{figure}
\centering
\scalebox{1}{\includegraphics{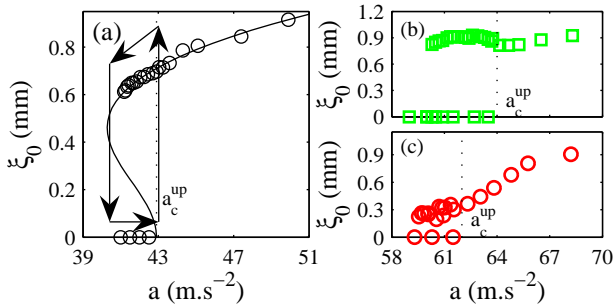}}
\caption{\label{fig3} (a) Amplitude $\xi_0$ of the surface waves versus driving acceleration $a$ in suspensions of {\it fd} viruses vibrated at $f=100$~Hz for three different concentrations: (a) $c=5.8$, (b) 11.3, and (c) 13.6~mg$/$mL. The black line in (a) is the best fit by Eq.~(\ref{eq.7}) with $a_0=42.9$~m.s$^{-2}$, $\eta_s=10^{-3}$~Pa.s, $\eta_1=5.9\eta_s$, $l_3=1.17$~mm, and $l_5=1.3$~mm. Dotted lines indicate $a_c^{\rm up}$ (see text).}
\end{figure}

The measurements of the surface wave amplitude $\xi_0$ shown in Fig.~\ref{fig3}(a) for the more dilute sample reveal that this hysteresis is associated with a discontinuous transition: above $a_c^{\rm up}$, the wave amplitude jumps directly to several hundreds of micrometers and drops abruptly to zero when $a$ is reduced by $\sim 4\%$. As seen in Fig.~\ref{fig3}(b) and (c), hysteresis is also observed for the sample close to the I--N transition and for the nematic sample. More precisely, the hysteresis gets more pronounced as one gets closer to the I--N transition, and abruptly decreases after the transition. This evolution with $c$ is reminiscent of that of the shear thinning properties of the sample~\cite{Dhont:2003a}. In these two concentrated samples, the hysteresis cycles also present more complex features and are less reproducible so that in the following we shall mostly focus on modelling the behaviour of the dilute sample. In any case, these effects are characteristic of a subcritical instability and can be attributed to the {\it shear-thinning} behaviour of our rodlike colloids. A simple argument relating shear-thinning to subcriticality is that a small perturbation of the surface may induce a shear rate that is large enough to significantly decrease the local viscosity, giving rise to a positive feedback on the perturbation and hence leading to finite wave amplitude at instability onset. As recalled above in the introduction, such an argument was the basis for a recent model by Epstein and Deegan \cite{Epstein:2010} who used piecewise linear fits of the viscosity vs. shear rate curve to account for high-amplitude strip waves in dilute wormlike micelle solutions. It was also shown that shear-thinning leads to a hysteretic, discontinuous transition.

\begin{figure}
\centering
\scalebox{1}{\includegraphics{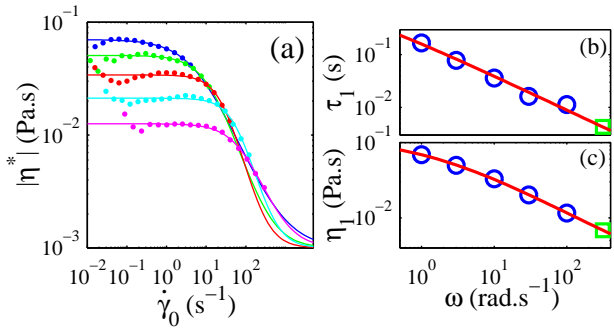}}
\caption{\label{fig2} (a) Modulus of the complex viscosity $|\eta^*|$ vs. amplitude $\dot\gamma_0$ of the oscillatory shear rate for the isotropic suspension at $c=5.8$~mg$/$mL. Colors refer to different oscillation pulsations: $\omega=1$ (\Blue{$\bullet$}), $\omega=3$ (\Green{$\bullet$}), $\omega=10$ (\Red{$\bullet$}), $\omega=30$ (\Cyan{$\bullet$}), and $\omega=100$~rad.s$^{-1}$ (\Magenta{$\bullet$}). The corresponding continuous lines are the best fits by Eq.~(\ref{eq.fit_rheo}) with $\eta_s=10^{-3}$~Pa.s. (b) Characteristic time $\tau_1$ and (c) viscosity $\eta_1$ extracted from the fits by Eq.~(\ref{eq.fit_rheo}) versus $\omega$ (\Blue{$\circ$}) and their extrapolations at $50$~Hz (\Green{$\square$}) from quadratic fits in logarithmic scales (red lines). 
}
\end{figure}

However, the model of Ref.~\cite{Epstein:2010} was based on steady-shear rheology whereas the flow induced by Faraday waves is {\it oscillatory}. Therefore, in the present study, we propose a simple alternative approach based on rheological measurements under large amplitude oscillatory shear. Figure~\ref{fig2} shows data recorded on the dilute sample using a standard rheometer (TA Instruments AR2000N). It is seen in Fig.~\ref{fig2}(a) that, whatever the oscillation pulsation $\omega$, the modulus of the complex viscosity $\eta^*$ as a function of the amplitude of the oscillatory shear rate $\dot\gamma_0$ is well fitted by the Cross model \cite{Cross:1965}:
\begin{equation}
\label{eq.fit_rheo}
|\eta^*|(\dot\gamma_0)=\frac{\eta_1}{1+(\dot\gamma_0 \tau_1)^m}+\eta_s\,,
\end{equation}
where $\tau_1$ is a characteristic time, $m$ an exponent, and $\eta_s$ the solvent viscosity taken equal to the water viscosity ($\eta_s=10^{-3}$~Pa.s). The coefficient $\eta_1$ can be interpreted as $\eta_1=\eta_0-\eta_s$, where $\eta_0$ is the modulus of the zero-shear complex viscosity at $\omega$. For all fits, we get $m=1.5\pm 0.1$ and the dependence of $\tau_1$ and $\eta_1$ on $\omega$ is shown in Fig.~\ref{fig2}(b) and (c) respectively. Since most classical rheometers are limited to oscillation frequencies below about 10~Hz, we use quadratic fits in logarithmic scales to extrapolate both $\tau_1$ and $\eta_1$ to 50~Hz, {\it i.e.} to half the driving frequency used in the Faraday experiment [see red lines and green symbols in Fig.~\ref{fig2}(b) and (c)]. Finally, in order to keep the analytical expressions tractable while including the above shear-thinning rheology, we shall thereafter replace the exponent $m=1.5$ by $m=2$ in Eq.~(\ref{eq.fit_rheo}). We do not expect this approximation to change the results qualitatively but it should be noted that it sharpens the decrease of the viscosity around the characteristic shear rate $\dot\gamma_1=1/\tau_1=235$~s$^{-1}$ extrapolated at 50~Hz.

Coming back to Faraday waves, the oscillating shear rate at the surface reads $\dot\gamma = \xi_0 k_c\omega e^{i\omega t/2}/2$, where $k_c$ is the critical wave number of the surface pattern (obtained by considering half a mode in the $y$ direction \cite{Douady:1990}: $k_c=\sqrt{k^2+\pi^2/l^2}$. For the dilute sample, amplitude measurements give $\xi_0\simeq6.10^{-4}$~m and $k_c\simeq1380$~m$^{-1}$ at $f=100$~Hz, yielding $|\dot\gamma |\simeq 260$~s$^{-1}$. This shear rate is comparable to the characteristic shear rate $\dot\gamma_1$ [see Fig.~\ref{fig2}(a)] so that a strong effect of Faraday waves on the fluid viscosity is expected. At this stage, a model based on the numerical integration of the Mathieu equation could be developed following Ref.~\cite{Epstein:2010}. Here, however, we propose to stick with analytical expressions by considering our shear-thinning fluid as an {\it effective Newtonian fluid}. More precisely, provided the fluid layer is deep enough ($k_c h\gg 1$) and the viscosity is low enough ($2\eta k_c^2/\omega\ll 1$), which is indeed the case in our experiments where $k_c h\simeq13$ and $2\eta k_c^2/\omega\simeq 0.06$, the local shear rate $\dot\gamma (z,t)= \dot\gamma_0 (z)e^{i\omega t/2}$ is known to decrease exponentially with the depth $z$: 
\begin{equation}
\label{eq.1}
\dot\gamma_0(z)= \frac{\xi_0 k_c\omega}{2} e^{zk_c}\,,
\end{equation}
where $-h\le z\le 0$ by convention~\cite{Kumar:1999}. Using Eq.~(\ref{eq.fit_rheo}) with $m=2$ yields the $z$-dependence of the viscosity: 
\begin{equation}
\label{eq.2}
|\eta^*|(z)= \frac{\eta_1}{1+\frac{\dot\gamma_0 (z)^2}{\dot\gamma_1^2}}+\eta_s\,.
\end{equation}
Next, we define $\eta_{\rm eff}$, the viscosity of a Newtonian fluid which dissipates the same energy as our colloidal suspension for a given surface wave amplitude $\xi_0$, as:
\begin{equation}
\label{eq.3}
\eta_{\rm eff}= \frac{\int_{-\infty }^0 \eta (z) \dot\gamma (z)^2 \,{\rm d}z}{\int_{-\infty }^0  \dot\gamma (z)^2 \,{\rm d}z}\,,
\end{equation}
where $k_c h\gg 1$ was used to extend the integration from $-h$ to $-\infty$. Introducing the characteristic amplitude $\xi_c=2\dot\gamma_1/(\omega k_c)$, Eqs.~(\ref{eq.2}) and (\ref{eq.3}) readily lead to:
\begin{equation}
\label{eq.3bis}
\eta_{\rm eff}= \eta_1 \frac{\xi_c^2}{\xi_0^2} \ln \left( 1+ \frac{\xi_0^2}{\xi_c^2} \right) +\eta_s \,.
\end{equation}
Finally, for Newtonian fluids, it is well known that the amplitude of the surface wave should evolve with time as~\cite{Douady:1990}: 
\begin{equation}
\label{eq.4}
\tau \,\frac{\partial \xi_0}{\partial t}=\frac{a-a_c}{a_c}\,\xi_0 -\frac{\xi_0^3}{l_3^2} -\frac{\xi_0^5}{l_5^4},
\end{equation}
where $a_c$ is the critical acceleration, $(a-a_c)/(a_c\tau)$ is the growth rate, and $l_3$ and $l_5$ are characteristic lengths linked to the dissipation process. For Newtonian fluids of low viscosity, it was also shown that the critical acceleration is proportional to the viscosity~\cite{Douady:1990}: $a_c\propto \eta_{\rm eff}$. Therefore, at equilibrium ($\partial/\partial t=0$), combining Eqs.~(\ref{eq.3}) and (\ref{eq.4}) yields the relationship between $a$ and $\xi_0$:
\begin{equation}
\label{eq.7}
a  = \frac{a_0}{\eta_1+\eta_s} \left(\eta_1\,\frac{\xi_c^2}{\xi_0^2} \ln \left( 1+ \frac{\xi_0^2}{\xi_c^2} \right) +\eta_s\right) \left( 1+\frac{\xi_0^2}{l_3^2}+\frac{\xi_0^4}{l_5^4} \right) \,,
\end{equation}
where $a_0$ is the critical acceleration of the fluid at rest ({\it i.e.} for $\xi_0=0$) and corresponds to the acceleration $a_c^{\rm up}$ measured experimentally. When $\xi_0\rightarrow 0$, the second-order expansion of the previous expression in $\xi_0^2$ reads: 
\begin{equation}
\label{eq.8}
a = a_0  \left( 1+\frac{\xi_0^2}{l_3^2}-\frac{\eta_1}{\eta_1+\eta_s}\,\frac{\xi_0^2}{2\xi_c^2} +\frac{\xi_0^4}{l_5^4}+\frac{\eta_1}{\eta_1+\eta_s}\,\frac{\xi_0^4}{3 \xi_c^4}\right) \,,
\end{equation}
so that the instability is supercritical for $l_3<\xi_c \sqrt{2(1+\eta_s/\eta_1)}$ and subcritical otherwise. In other words, when $\xi_c$ is small enough or equivalently when shear-thinning sets in at a low enough shear rate $\dot\gamma_1$, subcriticality and hysteresis are expected in the Faraday instability. Figure~\ref{fig3}(a) shows that Eq.~(\ref{eq.7}) provides a good fit of the experimental $a(\xi_0)$ data with only two free parameters, $l_3$ and $l_5$, since $\eta_s$ is known, $a_0$ is measured independently as $a_c^{\rm up}$, and $\eta_1$ is extracted from Fig.~\ref{fig2}(c). Still, the predicted curve goes to zero for an acceleration smaller than the experimental $a_c^{\rm down}$. We believe that the main reasons for this discrepancy are the very simple form of Eqs.~(\ref{eq.2}) and (\ref{eq.3}) along with possible dissipation at the walls induced by the presence of a meniscus, which is not taken into account in our simple model. We shall come back to the two more concentrated samples in the next section.

\section{Birefringence measurements}

\begin{figure}
\centering
 \scalebox{1}{\includegraphics{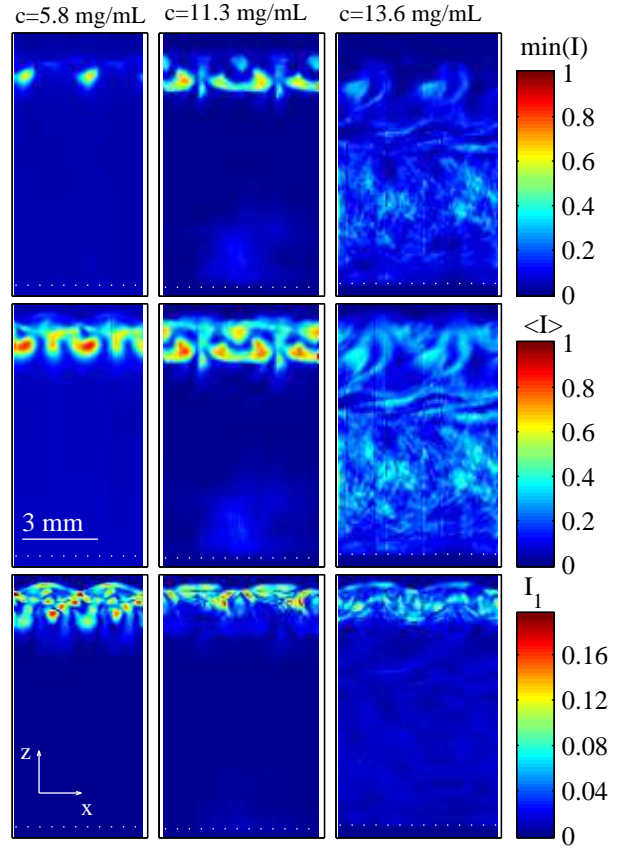}}
 \caption{\label{fig1} From top to bottom: minimum of the transmitted intensity $\min(I)$, time-averaged intensity $\left\langle I\right\rangle$, and amplitude $I_1$ of the first harmonic of the intensity recorded in suspensions of {\it fd} viruses vibrated at $f=100$~Hz for three different concentrations: $c=5.8$, 11.3, and 13.6~mg$/$mL from left to right. The white dotted lines indicate the bottom of the cell. All movies were taken at $a=a_c^{\rm up}$. Intensities are normalized by the maximum intensity of a given movie.$^\dag$}
\end{figure}

\begin{figure}
\centering
 \scalebox{1}{\includegraphics{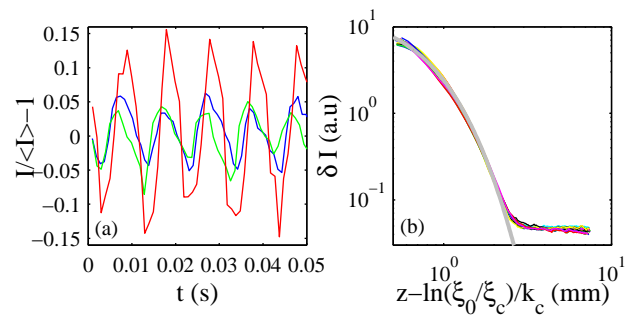}}
 \caption{\label{fig4}
(a) Normalized transmitted intensity $I/\langle I\rangle-1$ versus time $t$ recorded at an intensity antinode in suspensions of {\it fd} viruses vibrated at $f=100$~Hz for three different concentrations: $c=5.8$ (red), 11.3 (green), and 13.6~mg$/$mL (blue), and $a=1.1\,a_c^{\rm up}$. (b) $\delta I(z)$ versus $z-\ln(\xi_0/\xi_c)/k_c$ recorded in the more dilute suspension ($c=5.8$~mg$/$mL) at $100$~Hz and for different vibration amplitudes (see text). The grey line represents Eq.~(\ref{eq.9}) where $\dot\gamma_0(z)$ is given by Eq.~(\ref{eq.1}) and $\vert\eta^*\vert(z)$ by Eq.~(\ref{eq.2}). Except for the prefactor $\kappa$, all the parameters are inferred from the rheological measurements of Fig.~\ref{fig2}.}
\end{figure}

The above model provides a link between hysteresis and shear-thinning induced by the alignment of the rodlike colloids. This alignment is directly confirmed experimentally in Fig.~\ref{fig1} that reports the analysis of the transmitted intensity field $I(x,z,t)$ recorded over twenty driving periods with crossed polarizers:$^\dag$ the minimum intensity and the time-averaged intensity are presented in the first two rows while the last row shows the amplitude $I_1$ of the first harmonic of $I(x,z,t)$, {\it i.e.} of the component of $I$ that oscillates at $\omega$. In all cases, bright spots are observed near the surface that are evidence of a localized alignment of the {\it fd} viruses. More precisely, as long as the system remains far from a phase transition, the transmitted intensity field can be interpreted using the ``stress optical rule'' \cite{Decruppe:2003} according to which the birefringence $\Delta n$ is proportional to the shear stress $\sigma$, so that $I\propto \Delta n^2\propto \sigma^2$. Therefore, at least for the more dilute sample which is far from the I--N transition, the birefringence intensity should oscillate at $\omega$ and the amplitude of the first harmonic should read:
\begin{equation}
\label{eq.9}
I_1(x,z)=\kappa \sin^2(kx)\, \vert\eta^*(z)\vert^2\,\dot\gamma_0^2(z)\,,
\end{equation}
where $\kappa$ is a constant that depends on the exact geometry of the experiment, the optical set up, and the optical properties of the sample. Figure~\ref{fig4}(a) shows that the transmitted intensity indeed oscillates at $\omega$. Moreover, for the dilute sample, $I_1(x,z)$ can be predicted from the model developed above by inserting Eqs.~(\ref{eq.1}) and (\ref{eq.2}) into Eq.~(\ref{eq.9}). We restrain ourselves to the $z$-dependence of the transmitted intensity by calculating the average over $x$ of the temporal standard deviation of the transmitted intensity $I(x,z,t)$: $\delta I(z)=<\sqrt{<(I(x,z,t)-<I(x,z,t)>_t)^2>_t}>_x$ (where $<>_i$ denotes the average over $i$), which should be proportional to $\vert\eta^*(z)\vert^2 \dot\gamma_0^2(z)$. As seen in Fig.~\ref{fig4}(b), this approach provides a good description of $\delta I(z)$ with the constant $\kappa$ as the only free parameter. Also, when plotted versus $z-\ln(\xi_0/\xi_c)/k_c$, all curves $\delta I(z)$ measured for different vibration amplitudes collapse on the same curve. Incidentally, this also proves that $\kappa$ does not depend on $\xi_0$. More importantly, the results shown in Figs.~\ref{fig1}(left) and \ref{fig4} for the more dilute sample are consistent with Eq.~(\ref{eq.9}) and accredit a scenario of shear-induced alignment in which the rodlike colloids simply orient periodically according to the shear-rate field generated by the surface wave pattern.

For samples close to or above the I--N transition, however, the stress optical rule breaks down as the order parameter saturates and the above approach is no longer justified. Still, interesting new effects show up in Fig.~\ref{fig1}(middle) for the sample at $c=11.3$~mg$/$mL. There, it is seen that the minimum intensity is nonzero over a thin layer below the surface covering almost the whole width of the sample. This layer also corresponds to maxima in $\langle I\rangle$ while it does not show in $I_1$. This means that the surface waves have generated a {\it permanent} birefringence pattern below the surface with the same wavelength.$^\dag$ In other words, just below the I--N transition, the sample keeps the imprint of the surface wave pattern. The fact that the continuous component of the transmitted intensity becomes larger for larger concentrations is also reflected in the smaller oscillations of the relative amplitude reported in Fig.~\ref{fig4}(a): in the more concentrated samples, the oscillating component represents less than 5~\% of the average intensity, about three times smaller than for the dilute sample.

As for the most concentrated sample at $c=13.6$~mg$/$mL, it is easily checked from $\langle I\rangle$ in Fig.~\ref{fig1}(middle) that this nematic phase presents some significant birefringence everywhere in the sample. Moreover, although the effect is less spectacular than for the sample at $c=11.3$~mg$/$mL due to this natural birefringence, a continuous pattern is also generated that reflects the surface wave pattern. We interpret the build-up of a continuous birefringence field as a trade-off between rotational relaxation of the colloids and shear-induced alignment. In steady state, the rodlike viruses align locally due to shear, but close to the I--N transition, their large relaxation time prevents them from reorienting between two oscillations so that they remain aligned. This qualitative argument, which is developed more formally below, may also explain why the hysteresis cycles of the concentrated suspensions could not be properly accounted for [see Fig.~\ref{fig3}(b) and (c)]: since it takes longer for the fluid to relax after alignment, our experimental determination of the critical acceleration may not be accurate enough leading to some irreproducibility in $a_c$. 

\section{Numerical analysis}

A basic understanding of the virus alignment induced by Faraday waves can be obtained through numerical analysis following the calculation by Dhont and Briels for rigid cylindrical particles of length $L$ and diameter $D$ \cite{Dhont:2003a,Dhont:2003b}. In short, this approach yields the temporal evolution of the orientational order parameter tensor $\mathbf{S}$ based on the shear-rate field. One result is that, in the isotropic phase ($L\phi/D<4$), the linear complex viscosity $\eta^*(\omega)$ follows the Oldroyd-B model, where the characteristic variables depend on the solvent viscosity $\eta_s$, the volume fraction of rodlike particles $\phi$, their aspect ratio $L/D$, and their rotational diffusion coefficient at infinite dilution $D_r$, according to:
\begin{eqnarray}
\label{eq_oldB}\eta^*(\omega)=\tilde{\eta}_s+\frac{\tilde{\eta}_1}{1+i\omega\tau}\,,\\
\tilde{\eta}_s=\eta_s\left(1+\frac{\alpha\phi}{4}\right)\,,\\
\tilde{\eta}_1=\frac{3\eta_s \alpha \phi}{4}\,,\\
\tau=\left(6D_r\left(1-\frac{L\phi}{5D}\right)\right)^{-1}\,,\\
{\rm with~} \alpha=\frac{2}{5}\left(\frac{2L}{3D}\right)^2\frac{1}{\ln (L/D)}\,.
\end{eqnarray}
It should be noted that the above expression is only valid in the isotropic state and that extra care should be taken in the nematic state by fitting the calculated viscosity. Under this model, the sample is isotropic for $L\phi/D<40/9$ and nematic for $L\phi/D>40/9$. Moreover no direct comparison can be made between rheological results from this model and experimental ones as this model neglects the flexibility of the rodlike colloids which becomes important at high frequency due to Rouse modes, while at low frequency the particle diffusion is neglected~\cite{Morse:1998}. Still this calculation has proven useful to predict the alignment of rigid rods under shear~\cite{Lonetti:2008}, which we investigate here.

\begin{figure}
\centering
\scalebox{1}{\includegraphics{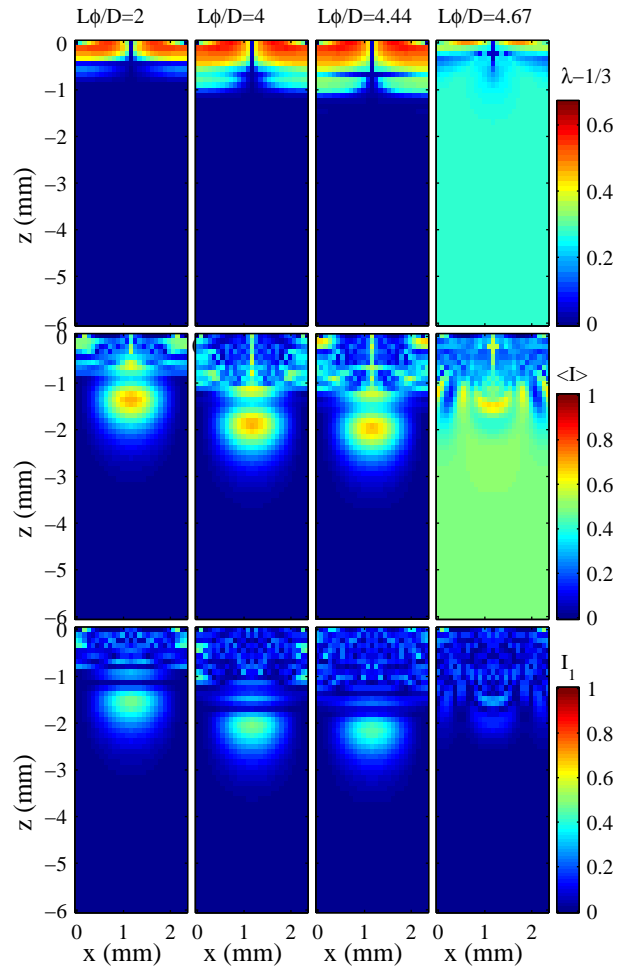}}
\caption{\label{fig_int_num5} From top to bottom: minimum of the order parameter minus its isotropic value $\lambda-1/3$, time-averaged birefringence intensity $\left\langle I\right\rangle$, and amplitude $I_1$ of the first harmonic of the intensity calculated for three concentrations approaching the I--N transition $L\phi/D=2$, $4$, and $4.44$, and a concentration in the nematic phase $L\phi/D=4.67$.}
\end{figure}

In the limit of small amplitude (linear approximation) in an horizontally unbounded cell, Ref.~\cite{Kumar:1999} gives the numerical expression of the two-dimensional velocity field $\mathbf{v}(x,z)$ regime knowing the complex viscosity of the sample. Using Eq.~(\ref{eq_oldB}), we first calculate the $\mathbf{v}(x,z)$ in the linear regime. We then deduce the shear rate field from which we calculate the local order parameter tensor $\mathbf{S}$ following Refs.~\cite{Dhont:2003a,Dhont:2003b} with only one additional input, namely the amplitude $\xi_0$ of the surface wave. 

We also compute the birefringence intensity $I$ from the order parameter $\mathbf{S}$:
\begin{eqnarray}
I=I_0\,\frac{\sin(2(\beta-\theta))^2(1-\cos(2\pi l \Delta\nu / \lambda))}{4}\,,\\
{\rm with~} \Delta\nu=\frac{4\pi\nu}{18}\,\frac{(\bar{n}^2+2)^2}{\bar{n}}\,\Delta\alpha\, \sqrt{(S_{xx}-S_{zz})^2+4 S_{xz}^2}\,,\label{eqnu}\\
{\rm and~} \theta=\frac{1}{2}\arctan\left(\frac{2S_{xz}}{S_{xx}-S_{zz}}\right)\,,
\end{eqnarray}
where the input parameters are the orientation of the pair of crossed polarizers $\beta$, the sample thickness $l$, the incident light wavelength $\lambda=575$~nm chosen as yellow (although we used white light experimentally), the normalized concentration $\nu= 4\phi/(\pi D^2 L)$, the mean refractive index approximated as the water index $\bar{n}=1.33$, and the polarizability anisotropy $\Delta\alpha=1.5.10^{-25}$~m$^{-3}$ deduced from Ref.~\cite{Purdy:2003}. For the calculations of Fig.~\ref{fig_int_num5}, we take $\beta=45^{\circ}$, $l=6$~mm, and $\xi_0=600$~$\mu$m consistently with the experimental values.

From these numerical calculations, we extract the minimum of the order parameter $\lambda$, the mean intensity $\langle I\rangle$, and the amplitude of the first harmonic $I_1$ of the intensity field (respectively shown from top to bottom in Fig.~\ref{fig_int_num5}) for three isotropic suspensions ($L/D \phi=2$, 4, and 4.44) and one in the nematic phase ($L/D\phi=$4.67). The minimum of the order parameter $\lambda(x,z)$ (first row of Fig.~\ref{fig_int_num5}) is calculated as the minimum during an oscillation of the maximum eigenvalue of $\mathbf{S}(x,z,t)$ and should be equal to $1/3$ for isotropic systems and greater than $1/3$ in the case of an out-of-equilibrium I--N transition. Close to the surface, a band is present where $\lambda > 1/3$ with a periodicity $2 k$. This is a clear indication of a phase transition induced by the surface waves and is similar to the bright zones of the minimum of intensity seen in Fig.~\ref{fig1}. The various features of the experimental birefringence pattern are qualitatively well captured. Especially, the width of the aligned band increases when approaching the I--N transition. This is a signature of the slowing down of the system at the I--N transition: as the relaxation time increases, the critical shear rate needed to align the suspension decreases.

Although the qualitative features of Fig.~\ref{fig1} are well reproduced by the numerical calculations, significant discrepancies are observed, most importantly the presence of higher order of birefringence ($2\pi l \Delta\nu / \lambda > \pi$) resulting in a secondary maximum in the computed intensity field that is absent in the experiment. This intensity peak marks the end of higher order birefringence. At the position where this maximum occurs, one has $\lambda=1/3$. Therefore, at this point, the sample is not permanently aligned. Since all the prefactors entering $\Delta\nu$ are well known [see Eq.~(\ref{eqnu})], we conclude that the order parameter $\mathbf{S}$ is different in the experiment and in the calculation. Many reasons can lead to this discrepancy. As already mentioned, the rheology deduced from Refs.~\cite{Dhont:2003a,Dhont:2003b} differs from the experimental rheological behaviour which leads to different velocity profiles. The calculation also supposes that the deformations remain small (linear hypothesis). As such, we should have $\xi_0 k \ll 1$ which is not the case in our experiments where $\xi_0 k_c\simeq 1$. Thus, in order to reach quantitative agreement, nonlinear terms related to the finite amplitude of the surface waves should be taken into account while calculating the velocity field.
Finally, the presence of a meniscus at the edge of the cell can damp the velocity. Still, the basic calculations presented in Fig.~\ref{fig_int_num5} strongly support our interpretation of the experimental observations in terms of a localized isotropic-to-nematic transition triggered by the Faraday instability.

\section{Conclusion}

The present study has unveiled a strong effect of the Faraday instability on suspensions of rodlike colloidal particles. By measuring the surface height profile, we have shown that the transition to parametrically excited surface waves displays discontinuous, hysteretic features. We first linked this subcritical behaviour to the shear-thinning properties of our colloidal suspensions thanks to a phenomenological model based on rheological data under large amplitude oscillatory shear. We have further provided evidence that Faraday waves induce local nematic ordering of the rodlike colloids through birefringence measurements. While local alignment simply follows the surface oscillations for dilute, isotropic suspensions, permanent nematic patches are generated by surface wave in samples close to the isotropic-to-nematic transition. Above the I--N transition nematic domains align in the flow direction. Such a strong coupling between the fluid microstructure and a hydrodynamic instability was confirmed by numerical computations. These results clearly differ from what was already observed in semidilute wormlike micelles~\cite{ Ballesta:2005} where the patterning of the birefringence intensity came from stretching of the micellar network without any phase transition.

Although more experiments are needed to fully understand the behaviour of rodlike colloids in complex shear fields, our results show that localized isotropic-to-nematic phase transitions can be induced in temporally stable structures with the same spatial organization as the surface instability pattern. This may open new paths of research, both experimental and theoretical, as well as potential applications, in which a complex system is structured at a mechanically or hydrodynamically controlled length scale intermediate between the mesoscopic size of the microstructure and the macroscopic size of the container.

\begin{acknowledgments}
The experiments were performed while the authors were at Centre de Recherche Paul Pascal (CRPP). We wish to thank the ``Cellule Instrumentation'' of CRPP for technical advice and design of the experiment as well as E.~Grelet, S.~Lerouge, and J.~Dhont for enlightening discussions.
\end{acknowledgments}


\begin{thebibliography}{0}

\bibitem{Faraday:1831}
M.~Faraday, {\it Philos. Trans. R. Soc. Lond.}, 1831, {\bf 52}, 319--340.

\bibitem{Kumar:1999}
S.~Kumar, {\it Phys. Fluids}, 1999, {\bf 11}, 1970--1981;
{\it Phys. Rev. E}, 2002, {\bf 65}, 026305.

\bibitem{Raynal:1999}
F.~Raynal, S.~Kumar, and S.~Fauve, {\it Eur. Phys. J. B}, 1999, {\bf 9}, 175--178.

\bibitem{Wagner:1999}
C.~Wagner, H.~W.~M{\"u}ller, and K.~Knorr, {\it Phys. Rev. Lett.}, 1999, {\bf 83}, 308--311.

\bibitem{Lioubashevski:1999}
O. Lioubashevski, Y. Hamiel, A. Agnon, Z. Reches, and J. Fineberg, {\it Phys. Rev. Lett.}, 1999, {\bf 83}, 3190--3193.

\bibitem{Merkt:2004}
F. S. Merkt, D. I. Goldman, E. C. Rericha, and H. L. Swinney, {\it Phys. Rev. Lett.}, 2004, {\bf 92}, 184501.

\bibitem{Huber:2005}
P. Huber, V. P. Soprunyuk, J. P. Embs, C. Wagner, M. Deutsch, and S. Kumar, {\it Phys. Rev. Lett.}, 2005, {\bf 94}, 184504.

\bibitem{Ballesta:2005}
P. Ballesta, and S. Manneville, {\it Phys. Rev. E}, 2005, {\bf 71}, 026308; {\it J. Non-Newtonian Fluid Mech.}, 2007, {\bf 147}, 23--34. 

\bibitem{Kityk:2006}
A. V. Kityk, and C. Wagner, {\it Europhys. Lett.}, 2006, {\bf 75}, 441--447.  

\bibitem{Ballesta:2006}
P. Ballesta, and S. Manneville, {\it Europhys. Lett.}, 2006, {\bf 76}, 429--435.

\bibitem{Cabeza:2007}
C. Cabeza and M. Rosen, {\it Int. J. Bifurcation Chaos}, 2007, {\bf 17}, 1599--1607.

\bibitem{Epstein:2010}
T. Epstein and R. D. Deegan, {\it Phys. Rev. E}, 2010, {\bf 81}, 066310.

\bibitem{Hernandez:2010}
M. Hern{\'a}ndez-Contreras, {\it J. Phys.: Condens. Matter}, 2010, {\bf 22}, 035106.

\bibitem{Lapointe:1973}
J. Lapointe and D. A. Marvin, {\it Mol. Cryst. Liquid Cryst.}, 1973, {\bf 19}, 269--278.

\bibitem{Dogic:1997}
Z. Dogic and S. Fraden,
{\it Phys. Rev. Lett.}, 1997, {\bf 78}, 2417.

\bibitem{Dogic:2000}
Z. Dogic and S. Fraden, {\it Langmuir}, 2000, {\bf 16}, 7820--7824.

\bibitem{Graf:1993}
C. Graf, H. Kramer, M. Deggelmann, M. Hagenb\"uchle, C. Johner, C. Martin, and R. Weber, {\it J. Chem. Phys.}, 1993, {\bf 98}, 4920--4928.

\bibitem{Lettinga:2004}
M. P. Lettinga and J. K. G. Dhont, {\it J. Phys.: Condens. Matter}, 2004, {\bf 16}, S3929--S3939.

\bibitem{Dhont:2003a}
J.~K.~G. Dhont and W.~J.~Briels, {\it Colloid Surface A}, 2003, {\bf 213}, 131--156.

\bibitem{Lonetti:2008}
B. Lonetti, J. Kohlbrecher, L. Willner, J. K. G. Dhont, and M. P. Lettinga, {\it J. Phys.: Condens. Matter}, 2008, {\bf 20}, 404207.

\bibitem{Cross:1965}
M. M. Cross, {\it J. Colloid Sci.}, 1965, {\bf 20}, 417--437.

\bibitem{Douady:1990}
S. Douady,
{\it J. Fluid. Mech.}, 1990, {\bf 221}, 383--409.

\bibitem{Decruppe:2003}
J.~P.~Decruppe and A.~Ponton,
{\it Eur. Phys. J. E}, 2003, {\bf 10}, 201--208.

\bibitem{Dhont:2003b}
J.~K.~G. Dhont and W.~J.~Briels, {\it J. Chem. Phys.}, 2003, {\bf 118}, 1466--1478.

\bibitem{Morse:1998}
D. C. Morse, {\it Macromolecules}, 1998, {\bf 31}, 7044--7067.

\bibitem{Purdy:2003}
K.~R.~Purdy, Z. Dogic, S. Fraden, A. R\"uhm, L. Lurio, and S. G. J. Mochrie, {\it Phys. Rev. E}, 2003, {\bf 67}, 031708.





\end{thebibliography}
\end{document}